\documentclass[11pt]{article}

\usepackage{amssymb}
\usepackage{amsmath}
\usepackage{amsthm}
\usepackage[dvips]{graphics,color}
\usepackage{epsfig}
\usepackage{a4wide}
\def\var{\text{Var}}

\def\U{\mathcal{U}}

\begin{document}

\title{Forecasting with time-varying vector autoregressive models}
\author{K. Triantafyllopoulos\footnote{Department of Probability and Statistics,
Hicks Building, University of Sheffield, Sheffield S3 7RH, UK,
Email: {\tt k.triantafyllopoulos@sheffield.ac.uk}, Tel: +44 114 222
3741, Fax: +44 114 222 3759.}}

\date{\today}

\maketitle

\begin{abstract}

The purpose of this paper is to propose a time-varying vector
autoregressive model (TV-VAR) for forecasting multivariate time
series. The model is casted into a state-space form that allows
flexible description and analysis. The volatility covariance matrix
of the time series is modelled via inverted Wishart and singular
multivariate beta distributions allowing a fully conjugate Bayesian
inference. Model performance and model comparison is done via the
likelihood function, sequential Bayes factors, the mean of squared
standardized forecast errors, the mean of absolute forecast errors
(known also as mean absolute deviation), and the mean forecast
error. Bayes factors are also used in order to choose the
autoregressive order of the model. Multi-step forecasting is
discussed in detail and a flexible formula is proposed to
approximate the forecast function. Two examples, consisting of
bivariate data of IBM shares and of foreign exchange (FX) rates for
8 currencies, illustrate the methods. For the IBM data we discuss
model performance and multi-step forecasting in some detail. For the
FX data we discuss sequential portfolio allocation; for both data
sets our empirical findings suggest that the TV-VAR models
outperform the widely used VAR models.

\textit{Some key words:} Bayesian forecasting, multivariate time
series, stochastic volatility, state space, foreign exchange rates,
portfolio allocation.

\end{abstract}

\section{Introduction}\label{intro}

Over the past 30 years there has been an emerging literature on
multivariate time series (Hamilton, 1995; Tsay, 2002; L\"utkepohl,
2005). Multivariate time series forecasting is required in many
financial applications, for example to enable optimal portfolio
allocation or to construct trading strategies over sectors of the
market, or exchange rates. In this direction vector autoregressive
models (Sims, 1980; Tiao and Tsay, 1989; Ni and Sun, 2003) are well
established as flexible and useful multivariate models. A
comprehensive treatment of these models can be found in L\"utkepohl
(2005).

A recent advance in time series analysis is the development of
autoregressive (AR) models, and more generally autoregressive moving
average models, with time-varying coefficients. Such models are
developed as in Kitagawa and Gersch (1985, 1996), Dahlhaus (1997),
West {\it et al.} (1999), Prado and Huerta (2002), Andrieu {\it et
al.} (2003), Lundbergh {\it et al.} (2003), Francq and Gautier
(2004), Moulines {\it et al.} (2005), Huerta and Prado (2006),
Abramovich {\it et al.} (2007), Triantafyllopoulos and Nason (2007),
and Zhu and Wu (2007). All these studies refer to univariate time
series; attempts to model vector time series with time-varying
autoregressive (TV-VAR) models include Jiang and Kitagawa (1993),
Sarantis (2006), Sato {\it et al.} (2007), and Triantafyllopoulos
(2007). Although such models are capable of capturing the
time-varying behaviour of time series data, it is desirable that
multivariate stochastic volatility is included in the model, in
particular for situations of financial time series forecasting. For
example Uhlig (1997) uses a vector autoregressive (VAR) model that
incorporates a stochastic volatility component, but its application
is relatively limited due to the fact that one needs to resort to
simulation-based estimation techniques (in particular Monte Carlo
simulation). In a similar direction Aguilar and West (2000) describe
the use of particle filters for portfolio allocation, but for
forecasting and in particular for sequential multi-step forecasting,
it is desirable to resort to analytic methods that are fast and
computationally more stable than their simulation-based
counterparts.

Suppose that the $p\times 1$ vector time series $\{y_t\}$, which is
observed at roughly equal intervals of time $t=1,\ldots,N$, is
generated by the autoregressive model
\begin{equation}\label{model1}
y_t=\phi_{0t}+\Phi_{1t}y_{t-1}+\cdots+\Phi_{dt}y_{t-d}+\epsilon_t,
\quad \epsilon_t\sim N_p(0,\Sigma_t), \quad t>d,
\end{equation}
where $\phi_{0t}$ is a $p\times 1$ vector, $\Phi_{jt}$ is a $p\times
p$ matrix $(j=1,\ldots,d)$, $d$ is a positive integer (known as the
lag order of the autoregression), $\Sigma_t$ is a $p\times p$
covariance matrix, the sequence of innovations $\{\epsilon_t\}$ is
independent and $\epsilon_t$ follows a $p$-variate Gaussian
distribution.

In VAR estimation, a common practice is to recast the model in VAR
form of order 1, known as the reduced form, see e.g. Tiao and Tsay
(1989). It is possible to extend this to our model (\ref{model1}),
and write
$$
\left[\begin{array}{c} y_t\\ y_{t-1} \\ \vdots \\
y_{t-d+1}\end{array}\right] = \left[\begin{array}{c} \phi_{0t}\\ 0 \\
\vdots \\ 0\end{array}\right] + \left[\begin{array}{ccccc} \Phi_{1t}
& \Phi_{2t} & \cdots & \Phi_{d-1,t} & \Phi_{dt} \\ I_p & 0 & \cdots & 0 & 0\\
\vdots & \vdots & \ddots & \vdots & \vdots \\ 0 & 0 & \cdots & I_p &
0\end{array}\right] \left[\begin{array}{c} y_{t-1} \\ y_{t-2} \\
\vdots \\ y_{t-d}\end{array}\right] + \left[\begin{array}{c}
\epsilon_t \\ 0 \\ \vdots \\ 0\end{array}\right],
$$
which can be expressed as $Y_t=\psi_{0t}+\Psi_t Y_{t-1}+\eta_t$,
where $Y_t'=[y_t'~y_{t-1}'~\cdots~y_{t-d+1}']$,
$\psi_{0t}'=[\phi_{0t}'~0'~\cdots~0']$, and
$\eta_t'=[\epsilon_t'~0'~\cdots~0']$. Now the model for $\{Y_t\}$ is
a TV-VAR model of order 1. Although it can be claimed that $Y_t$
possesses a more compact model as that of $y_t$, in principle it is
harder to derive multi-step forecasting equations for $Y_t$, as the
forecast horizon will be included in the dimension of $Y_t$ and this
can cause technical problems. Furthermore, we notice that $\eta_t$
has a singular covariance matrix (only $p$ elements of $\eta_t$ are
non-degenerate) and so if one wishes to apply a prior distribution
on the covariance matrix of $\eta_t$, this has to be a singular
multivariate distribution, e.g. a singular Wishart distribution. As
a result the above reduced formulation causes some technical
difficulties. These difficulties may be reasonably overcome when
$p$, the dimension of $y_t$, and $d$, the AR order, are not too
large, but if at least one of $p$ or $d$ are large, then the
advantage of employing the reduced form model, seems to be lost.

The aim of this paper is to suggest alternative state-space forms
which will enable the modeller to overcome many of the difficulties
mentioned above. In particular, this paper proposes Bayesian
estimation for (\ref{model1}) after it is put in an appropriate
state-space form. The proposed model includes a stochastic evolution
that can update the volatility covariance matrix from time $t-1$ to
$t$. This evolutionary law is supported by inverted Wishart and
singular multivariate beta distributions, following developments of
Uhlig (1994) and Triantafyllopoulos (2008). We discuss multi-step
forecasting for this model and we propose a simple formula that
approximates the true forecast function. The choice of the
autoregressive order is an important problem (see e.g. de Waele and
Broersen, 2003; or Abramovich {\it et al.}, 2007), which here is
resolved by using sequential Bayes factors as a means of model
selection. For model judgment and model comparison we discuss
several measures of goodness of fit (mean of squared standardized
forecast errors, mean absolute forecast error, and mean forecast
error) as well as a likelihood-based criterion and sequential Bayes
factors. Two examples, consisting of bivariate data of IBM shares
and foreign exchange (FX) rates of 8 currencies, illustrate the
methods. For the IBM data set we discuss in detail model performance
and multi-step forecasting; for the FX data we discuss a sequential
portfolio selection strategy. For both data sets the time-varying
model is found to outperform its time-invariant counterparts and as
a result we propose that TV-VAR models are superior to VAR models.

The rest of the paper is organized as follows. In the next section
we define the model, which estimation and forecasting is discussed
in Section \ref{estimation}. Sections \ref{example:ibm} and
\ref{example:fx} include the data analyses and the paper closes with
brief concluding comments.

\section{Model setting}\label{modelsetting}

Consider model (\ref{model1}), with $\phi_{0t}$ and $\Phi_{jt}$
generated by a random walk. This model postulates that
$\phi_{0t}-\phi_{0,t-1}\approx 0$ and $\Phi_{jt}-\Phi_{j,t-1}\approx
0$. Other evolutionary models may be considered, e.g. a Markovian
evolution, but here we focus our attention on the random walk model.

Write
$$
\Phi_t'=\left[\begin{array}{cccc} \phi_{0t} & \Phi_{1t} & \cdots &
\Phi_{dt}\end{array}\right],
$$
where $\Phi_t'$ denotes the transpose matrix of $\Phi_t$ so that
$\Phi_t$ is a $(dp+1)\times p$ matrix and $\Phi_t'$ is a $p\times
(dp+1)$ matrix.

Then we can express the above random walk as
\begin{equation}\label{model2}
\Phi_t=\Phi_{t-1}+\Omega_t,
\end{equation}
where we assume that the $(dp+1)\times p$ random matrix $\Omega_t$
follows a matrix-variate Gaussian distribution, i.e. $\Omega_t\sim
N_{(dp+1)\times p}(0,W_t,\Sigma_t)$. This implies that
$\textrm{vec}(\Omega_t)$ follows the $p(dp+1)$-variate Gaussian
distribution $\textrm{vec}(\Omega_t)\sim
N_{p(dp+1)}(0,\Sigma_t\otimes W_t)$, where $\textrm{vec}(.)$ denotes
the column stacking operator of a matrix and $\otimes$ denotes the
Kronecker product. Here the $(dp+1)\times (dp+1)$ covariance matrix
$W_t$ is assumed known, but later we discuss its specification using
several discount factors. It is also assumed that $\Omega_t$ is
independent of $\Omega_s$ $(t\neq s)$ and that for any $t,q$,
$\Omega_t$ is independent of $\epsilon_q$.

Model (\ref{model1}) can be written in state-space form as
\begin{equation}\label{model1a}
y_t' = [1~y_{t-1}'\cdots ~y_{t-d}']\left[\begin{array}{c} \phi_{0t}' \\ \Phi_{1t}' \\ \vdots \\
\Phi_{dt}'\end{array}\right] +\epsilon_t' = F_t'\Phi_t+\epsilon_t',
\end{equation}
where $\Phi_t$ is generated by evolution (\ref{model2}). We use the
notation $y^t$ for the information set up to time $t=1,\ldots,N$,
i.e. $y_t=(y_1,\ldots,y_t)$.

It remains to define an evolution for $\Sigma_t$. Here we use the
model of Triantafyllopoulos (2008), which adopts the multiplicative
evolutionary law of Uhlig (1994), based on the convolution of the
Wishart and the singular multivariate beta distribution. For the
purpose of this paper, we assume that $\Sigma_t$ is a symmetric
positive-definite matrix and we propose the following law
\begin{equation}\label{model3}
\Sigma_t^{-1}=k\U(\Sigma_{t-1}^{-1})'B_t\U(\Sigma_{t-1}^{-1}),
\end{equation}
where $\U(\Sigma_{t-1}^{-1})$ denotes the upper triangular matrix of
the Choleski decomposition of $\Sigma_{t-1}^{-1}$, $B_t$ follows
independently of $\Sigma_{t-1}^{-1}$ a singular multivariate beta
distribution with degrees of freedom $k_1=(n+p-1)/2$ and $k_2=1/2$,
written $B_t\sim B(k_1,k_2)$. The singularity of this beta
distribution is reflected on the fact that $2k_2=1<p-1$, and so the
matrix $I_p-B_t$ is singular. The singular multivariate beta
density, which is defined on the Stiefel manifold, replaces the
determinant of $I_p-B_t$ (which is zero) by the product of the
positive eigenvalues of that matrix; for more details on the beta
distribution the reader is referred to Uhlig (1994). Here we set
$n=1/(1-\beta)$, where $\beta$ is a discount factor $(0<\beta<1)$
and $k$ is defined as $k=(\beta(1-p)+p)/(\beta(2-p)+p-1)$. The
choice of the beta distribution and the choice of $k$ are done so
that $\Sigma_t^{-1}$ resembles a random walk type evolution, i.e.
$E(\Sigma_t^{-1}|y^{t-1})=E(\Sigma_{t-1}^{-1}|y^{t-1})$ and
$\var(\textrm{vecp}(\Sigma_t^{-1})|y^{t-1})\geq
\var(\textrm{vecp}(\Sigma_{t-1}^{-1})|y^{t-1})$, where $\var(.)$
denotes covariance matrix and $\textrm{vecp}(.)$ denotes the column
stacking operator of a covariance matrix. This simply says that
going from time $t-1$ to $t$, the expectation of the volatility
remains unchanged, and the respective covariance matrix of the
vectorized form of the volatility is increased.

For $t\geq d+1,\ldots,N$, the model consists of equations
(\ref{model2}), (\ref{model1a}) and (\ref{model3}), together with
the priors
\begin{equation}\label{priors}
\Phi_d\sim N_{(dp+1)\times p}N(m_d, P_d, \Sigma_d) \quad
\textrm{and} \quad \Sigma_d^{-1} \sim W_p(n+2p, S_d),
\end{equation}
the latter of which, denotes a Wishart distribution with $n+2p$
degrees of freedom and parameter matrix $S_d$. The quantities
$m_d,P_d,S_d$ are assumed known. A weakly informative prior
specification suggests $m_d=\textrm{initial belief of } \Phi_d$,
$P_d=1000I_{dp+1}$ and $S_d=I_p$. The covariance matrix $W_t$, which
is responsible for the shocks of $\Phi_t$, can be defined by using a
discount matrix $\Delta$, i.e.
$\var(\textrm{vec}(\Phi_t)|\Sigma_{t-1},y^{t-1})=\Sigma_{t-1}\otimes
R_t$, with $R_t=\Delta^{-1/2}P_{t-1}\Delta^{-1/2}=P_{t-1}+W_t$ and
$P_{t-1}$ being known at time $t-1$ so that
$\var(\textrm{vec}(\Phi_{t-1})|\Sigma_{t-1},y^{t-1})=\Sigma_{t-1}\otimes
P_{t-1}$. The discount matrix $\Delta$ is the diagonal matrix of
$dp+1$ discount factors $\delta_j$, i.e.
$\Delta=\textrm{diag}(\delta_1,\ldots,\delta_{dp+1})$. Similar
discount models are discussed in West and Harrison (1997). The
discount factors $\delta_1,\ldots,\delta_{dp+1}$ and $\beta$ can be
regarded as hyperparameters and they can be specified by considering
goodness of fit criteria.

We note that, although the above model is very general, subclasses
of this include interesting and applicable models. For example, if
$\Phi_{1t}=\cdots=\Phi_{dt}=0$, then we obtain a traditional
stochastic volatility model
$y_t=\phi_{0t}+\Sigma_t^{1/2}\varepsilon_t$ and $\varepsilon_t\sim
N_p(0,I_p)$ (Triantafyllopoulos, 2008). If $\phi_{0t}=\phi$,
$\Phi_{jt}=\Phi$ are all time-invariant, then we obtain a VAR model
with stochastic volatility (Uhlig, 1997). If $\Sigma_t=\Sigma$ is
time-invariant (this can be achieved by setting $\beta=1$, but we
need to replace $n$ by $n_t=n_{t-1}+1=n_0+t$), then we obtain the
model of Triantafyllopoulos (2007); such a model may be useful for
very short periods of time (when the volatility can be assumed
time-invariant), or for time series that do not exhibit a
heteroscedastic behaviour (such data may arise in environmental
studies).

\section{Estimation and forecasting}\label{estimation}

\subsection{Bayesian estimation}

Bayesian estimation follows from a generalization of the Kalman
filter, the details of which can be found in Triantafyllopoulos
(2008). Here we briefly describe the algorithm. Suppose that at time
$t-1$, the posterior distributions of $\Phi_{t-1}$, given
$\Sigma_{t-1}$, and $\Sigma_{t-1}$ are given by
$\Phi_{t-1}|\Sigma_{t-1},y^{t-1}\sim N_{(dp+1)\times
p}(m_{t-1},P_{t-1},\Sigma_{t-1})$ and $\Sigma_{t-1}|y^{t-1}\sim
IW_p(n+2p,S_{t-1})$ (here $IW(.)$ denotes the inverted Wishart
distribution so that $\Sigma_{t-1}^{-1}$ follows a Wishart
distribution). Moving on to time $t$ with information $y^{t-1}$, we
have $\Phi_t|\Sigma_{t-1},y^{t-1}\sim N_{(dp+1)\times
p}(m_{t-1},R_t,\Sigma_{t-1})$ and $\Sigma_t|y^{t-1}\sim IW_p(\beta
n+2p,k^{-1}S_{t-1})$, where $R_t$ and $k$ are defined in the
previous section. Then by observing $y_t$, with information $y^t$,
we update the posteriors at $t$ as $\Phi_t|\Sigma_t,y^t\sim
N_{(dp+1)\times p}(m_t,P_t,\Sigma_t)$ and $\Sigma_t|y^t\sim
IW_p(n+2p,S_t)$, with $m_t=m_{t-1}+K_te_t'$, $P_t=R_t-K_tK_t'/Q_t$,
$K_t=R_tF_t/Q_t$, $Q_t=F_t'R_tF_t+1$, $e_t=y_t-m_{t-1}'F_t$, and
$S_t=k^{-1}S_{t-1}+e_te_t'/Q_t$. $K_t$ is called the adaptive vector
(which plays a similar role as the Kalman gain in the standard
Kalman filter), and $e_t$ is the one-step ahead forecast error. At
time $t=d$, starting with the priors (\ref{priors}), the above
posterior distributions give a recursive algorithm for any
$t=d+1,\ldots,N$.

Given some data $y_1,\ldots,y_N$ the likelihood function of
$\Sigma_{d+1},\ldots,\Sigma_N$ can be derived by
$$
L(\Sigma_{d+1},\ldots,\Sigma_N;y^N)=\prod_{t=d+1}^N
p(y_t|\Sigma_t)p(\Sigma_t|\Sigma_{t-1}),
$$
where $p(y_t|\Sigma_t)$ is the density function of $y_t$ conditional
on $\Sigma_t$, and $p(\Sigma_t|\Sigma_{t-1})$ is the density
function of $\Sigma_t$, given $\Sigma_{t-1}$. From the Kalman filter
we have $y_t|\Sigma_t\sim N_p(m_t'F_t,\Sigma_t)$ and from the
singular multivariate beta density of model (\ref{model3}), we
obtain
$$
p(\Sigma_t|\Sigma_{t-1})=\pi^{-p/2} k^{-p(n-p)/2}
\frac{\Gamma((n+1)/2)}{\Gamma(n/2)}|L_t|^{-p/2}
|\Sigma_{t-1}|^{(n-p)/2} |\Sigma_t|^{-(n-p-1)/2},
$$
where $n$ and $k$ are defined in Section \ref{modelsetting}, and
$L_t$ is the diagonal matrix with elements the non-zero eigenvalues
of the matrix $I_p-k^{-1}(\U(\Sigma_{t-1}^{-1})')^{-1}\Sigma_t^{-1}
\U(\Sigma_{t-1}^{-1})$. This result is derived by the evolution of
the volatility (\ref{model3}) and the singular multivariate beta
density of $B_t$, details of which can be found in
Triantafyllopoulos (2008). Then the log-likelihood function of
$\Sigma_{d+1},\ldots,\Sigma_N$ is
\begin{eqnarray*}
\ell(\Sigma_{d+1},\ldots,\Sigma_N;y^N) &=&
c-\frac{1}{2}\sum_{t=d+1}^N
(y_t-m_{t-1}'F_t)'\Sigma_t^{-1}(y_t-m_{t-1}'F_t) + \frac{n-p}{2}
\sum_{t=d+1}^N \log |\Sigma_{t-1}| \\ && - \frac{n-p}{2}
\sum_{t=d+1}^N \log |\Sigma_t| - \frac{p}{2} \sum_{t=d+1}^N \log
|L_t|,
\end{eqnarray*}
where
$$
c=-\frac{Np}{2}\log(2\pi^2) -\frac{Np(n-p)}{2}\log k + N\log
\frac{\Gamma_p((n+1)/2)}{\Gamma_p(n/2)}.
$$
From $\Sigma_t|y^t\sim IW_p(n+2p,S_t)$, by replacing the posterior
mean of $\Sigma_t|y^t$ in $\ell(\Sigma_{d+1},\ldots,\Sigma_N;y^N)$,
we can evaluate the above log-likelihood function for the given data
$y^N=(y_1,\ldots,y_N)$.

\subsection{Stability of $P_t$}\label{section:pt}

In this section we show that if $\{y_t\}$ is a bounded sequence,
then the sequence $\{P_t\}$ is also bounded. This is an important
property because if $\{P_t\}$ were unbounded, then the estimates of
$\Phi_{it}$ and the forecasts of $y_t$ would be totally unstable,
possibly having infinite variances. In particular, given data
$y_1,\ldots,y_N$, we assume that $||y_t||<l$, where $\parallel
.\parallel$ denotes the Eucledian norm of a vector or matrix and $l$
is some positive number. This in principle means that by writing
$y_t=[y_{1t}~\cdots~y_{pt}]'$, all $y_{jt}$ are bounded scalar time
series.

Since $\{y_t\}$ is bounded, $\{F_t\}$ is also bounded and thus
$\{F_tF_t'\}$ is bounded. From the recursion $P_t=R_t-Q_tK_tK_t'$ we
have $P_t^{-1}=R_t^{-1}+F_tF_t'$ (for all $t>d$) and so we can write
$$
P_t^{-1}=\Delta^{(t-d)/2}P_d^{-1}\Delta^{(t-d)/2} + \sum_{i=0}^{t-d}
\Delta^{i/2}F_{t-i}F_{t-i}'\Delta^{i/2} \approx \sum_{i=0}^{t-d}
\Delta^{i/2}F_{t-i}F_{t-i}'\Delta^{i/2},
$$
by adopting the weakly informative prior $P_d^{-1}\approx 0$. Then
$$
\parallel P_t^{-1} \parallel \leq  \sum_{i=0}^{t-d} \parallel
\Delta^{i/2} \parallel^2 \parallel F_{t-i}F_{t-i}' \parallel  \leq M
\sum_{i=0}^{t-d} (\delta_1^i+\cdots +\delta_{dp+1}^i) \leq M
\sum_{j=1}^{dp+1} \frac{1}{1-\delta_j}<\infty,
$$
which shows that $\{P_t^{-1}\}$ is bounded. If $P_d^{-1}\neq 0$, one
needs to add a factor in the bound of $P_t$, which again shows that
$P_t^{-1}$ is bounded, except if $P_d=0$. Thus, for any prior
$P_d>0$, the sequence $\{P_t^{-1}\}$ is bounded. Since $P_t$ is
non-singular, for any $t>d$, it follows that the sequence $\{P_t\}$
is also bounded, for any prior $P_d>0$.

\subsection{Choice of order $d$}

The performance of the model is usually judged against its forecast
performance, which is best described by its forecast distribution.
Thus sequential Bayes factors, which for univariate time series are
described in West and Harrison (1997), can be used in order to
compare and contrast models which differ in the order of the AR.
Suppose we have two TV-VAR models, which have respectively lag
orders $d_1$ and $d_2$ and thus they consequently differ in the
design vectors $F_{1t}$ and $F_{2t}$, which are functions of $d_1$
and $d_2$. According to the above, the one-step forecast
distributions are Student $t$, i.e. $y_{t+1}|y^t,\textrm{model
}i\sim t_p(n,y_{it}(1),Q_{it}^*(1))$ and thus the Bayes factor of
model 1 against model 2 is defined by
$$
H_t(1)=\frac{p(y_{t+1}|y^t,\textrm{model
}1)}{p(y_{t+1}|y^t,\textrm{model
}2)}=\left(\frac{|Q_{2t}^*(1)|}{|Q_{1t}^*(1)|}\right)^{1/2}
\left(\frac{\beta n+e_{1,t+1}'(Q_{1t}^*(1))^{-1}e_{1,t+1}}{\beta
n+e_{2,t+1}'(Q_{2t}^*(1))^{-1}e_{2,t+1}}\right)^{-(\beta n+p)/2},
$$
where $y_{it}(1)$ is the one-step forecast mean of $y_{t+1}$ (under
model $i$), and $Q_{it}^*(1)$ is the scale matrix of the $t$
distribution (see also the next section). Comparison of $H_t(1)$
with 1 can indicate preference of model 1 (if $H_t(1)>1$) or
preference of model 2 (if $H_t(1)<1$), while if $H_t(1)=1$, the two
models are equivalent. Thus we choose the order that yields larger
values in the respective forecast function. The above formula for
the Bayes factor can also be used more generally for comparison of
two models that differ in the values of the hyperparameters, e.g. in
the discount factors. the advantage of the above scheme is that it
can be applied sequentially and thus indicate preference of a model
at each time point. This can lead to sequential model comparison and
in the context of this section one can consider schemes where the
order of the AR is time-varying, although this is not developed
further here.

\subsection{Multi-step forecasting}

Forecasting commences by considering the density of $y_{t+h}|y^t$,
for some positive integer $h$, known as the forecast horizon. Denote
with $y_t(h)$ the $h$-step forecast mean of $y_{t+h}$, i.e.
$y_t(h)=E(y_{t+h}|y^t)$ and with $Q_t(h)$ the respective $h$-step
forecast covariance matrix. Note that in the state-space formulation
(\ref{model1a}), $F_t$ can be stochastic, as it includes values of
$y_t$. We denote with $\widehat{F}_{t+h}$ the estimate of $F_{t+h}$
by replacing the unknown values of $y_{t+j}$ by the known $y_t(j)$,
for all $j\leq h$. Note that if $h>d$ we have
$$
\widehat{F}_{t+h}'=\left[\begin{array}{cccc} 1&y_t'(h-1)&\cdots
&y_t'(h-d)\end{array}\right]
$$
while if $h\leq d$ we get
$$
\widehat{F}_{t+h}'=\left[\begin{array}{ccccccc} 1&y_t'(h-1)&\cdots
&y_t'(1)&y_t'&\cdots &y_{t+h-d}'\end{array}\right]
$$

From the evolution (\ref{model2}) of $\Phi_t$ we obtain
$$
\Phi_{t+h}=\Phi_t+\sum_{i=1}^h \Omega_{t+i}.
$$
From the posterior distribution $\Phi_t|\Sigma_t,y^t$ we obtain
$\Phi_{t+h}|\Sigma_{t+h},y^t\sim N_{(dp+1)\times
p}(m_t,R_t(h),\Sigma_{t+h})$ and so from
$y_{t+h}'=F_{t+h}'\Phi_{t+h}+\epsilon_{t+h}'$, the $h$-step forecast
density of $y_{t+h}$ can be approximated as $y_{t+h}|y^t\sim
t_{p\times 1}(\beta n, m_t'\widehat{F}_{t+h},
(\widehat{F}_{t+h}'R_t(h)\widehat{F}_{t+h}+1)k^{-1}S_t)$, where
$R_t(h)=P_t+\sum_{i=1}^hW_{t+i}$ and $S_{t+h}=k^{-1}S_t$. Here
$t_{p\times 1}(.)$ denotes the $p$-variate Student $t$ distribution,
from which it follows
\begin{equation}\label{for}
y_t(h)=m_t'\widehat{F}_{t+h} \quad \textrm{and} \quad Q_t(h)=
\frac{(\widehat{F}_{t+h}'R_t(h)\widehat{F}_{t+h}+1)(1-\beta)k^{-1}}{3\beta
-2} S_t,
\end{equation}
where $\beta>2/3$. For $h=1$, the forecast mean vector $y_t(1)$ and
the forecast covariance matrix $Q_t(1)$ are exact, as
$F_{t+1}'=[1~y_t'~\cdots~y_{t+1-d}']$, which, given $y^t$ is not
stochastic.

We can define the $h$-step forecast errors as
$e_t(h)=y_{t+h}-y_t(h)$, which approximately follow a $t$
distribution, i.e. $e_t(h)|y^t\sim t_{p\times 1}(\beta
n,0,Q_t^*(h))$, where
$Q_t^*(h)=(\widehat{F}_{t+h}'R_t(h)\widehat{F}_{t+h}+1)k^{-1}S_t$.
We can then define the standardized $h$-step forecast errors as
$u_t(h)=(Q_t^*(h))^{-1/2}e_t(h)$, where $(Q_t^*(h))^{-1/2}$ denotes
the symmetric square root matrix of $(Q_t^*(h))^{-1}$, so that,
approximately, $u_t(h)|y^t\sim t_{p\times 1}(\beta n, 0, I_p)$, from
which we can derive $100\alpha\%$ quantiles and so we can build
credible bounds for $e_t(h)$. Finally, from $E(e_t(h)|y^t)=0$ and
$\var(e_t(h)|y^t)=Q_t(h)$, we can define standardized versions of
$e_t(h)$ as $v_t(h)=(Q_t(h))^{-1/2}e_t(h)$ so that $E(v_t(h)|y^t)=0$
and $\var(v_t(h)|y^t)=I_p$. Then, a measure of goodness of fit is
the mean of squared standardized $h$-step forecast errors, defined
as
$MSSE(h)=(N-d-h+1)^{-1}\sum_{t=d}^{N-h}[v_{1t}^2(h)~\cdots~v_{pt}^2(h)]'$,
which, if the model fit is good should be close to the vector
$[1~\cdots~1]'$. A second measure of goodness of fit is the mean
absolute forecast error, known also as the mean absolute deviation,
defined as $MAE(h)=(N-d-h+1)^{-1}\sum_{t=d}^{N-h}
[|e_{1t}(h)|~\cdots~|e_{pt}(h)|]'$, where
$e_t(h)=[e_{1t}(h)~\cdots~e_{pt}(h)]'$ and $|.|$ indicates absolute
value. Finally a third measure of goodness of fit is the mean
forecast error, defined by $ME(h)=(N-d-h+1)^{-1}\sum_{t=d}^{N-h}
e_t(h)$, which can be used to measure how biased are the forecasts.
the log-likelihood function (evaluated at the posterior mean of the
volatility) and the Bayes factors (see also the previous section)
can be used for model performance and model comparison.

For the volatility, from the evolution (\ref{model3}) we have
$E(\Sigma_{t+1}^{-1}|y^t)=E(\Sigma_t^{-1}|y^t)$ and so we can find
some symmetric random matrix $\Xi_{t+1}$ with zero mean so that
$\Sigma_{t+1}^{-1}=\Sigma_t^{-1}+\Xi_{t+1}$, defining a random walk
evolution for $\Sigma_t^{-1}$. This implies that
$\Sigma_{t+h}^{-1}=\Sigma_t^{-1}+\sum_{i=1}^h\Xi_{t+i}$ and one way
to support this evolution of $\Sigma_t^{-1}$ is by defining
$\Sigma_{t+h}^{-1}=k\U(\Sigma_t^{-1})'B_{t+h}\U(\Sigma_t^{-1})$,
where $B_{t+h}$ follows, independently of $\Sigma_t^{-1}$, a
singular multivariate beta distribution with degrees of freedom
$n/2$ and 1/2 respectively. Then, from the Wishart distribution of
$\Sigma_t^{-1}$, we have $\Sigma_{t+h}^{-1}\sim W_p(\beta
n+p-1,kS_t^{-1})$ and so $\Sigma_{t+h}|y^t\sim IW_p(\beta
n+2p,k^{-1}S_t)$. Then we can write
$$
S_t(h)=E(\Sigma_{t+h}|y^t)=E(\Sigma_{t+1}|y^t)=\frac{(1-\beta)k^{-1}}{3\beta-2}
S_t.
$$

\begin{figure}
 \epsfig{file=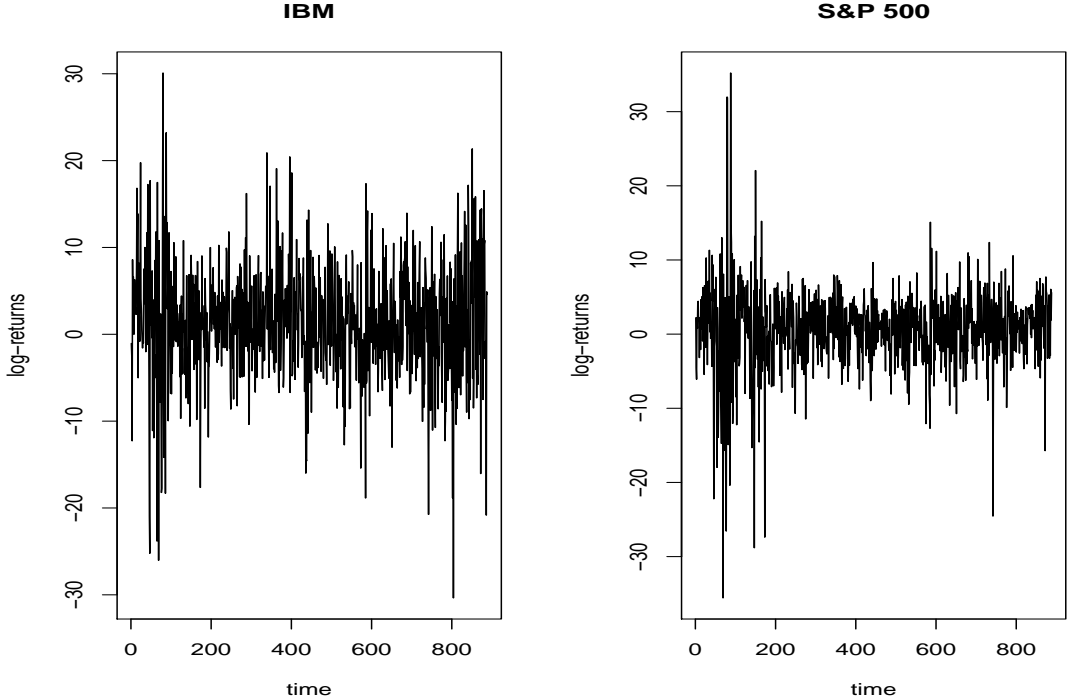, height=10cm, width=15cm}
 \caption{Log-returns for IBM shares and S\&P 500 index.}\label{fig1}
\end{figure}

\section{Forecasting IBM shares and S\&P 500 index
data}\label{example:ibm}

We consider bivariate time series data consisting of 888 log-returns
of IBM shares and of the S\&P 500 index. The data, which are plotted
in Figure \ref{fig1}, are collected in a daily frequency and they
are discussed in Tsay (2002, Chapter 9). Here we propose the use of
TV-VAR models with stochastic volatility in order to forecast time
series and the volatility. Table \ref{table1} shows the
log-likelihood function (evaluated at the posterior mean of
$\Sigma_i$), the mean of squared standardized one-step forecast
errors ($MSSE(1)$) and the mean absolute one-step forecast error
($MAE(1)$), for a variety of TV-VAR models. Here, for the evolution
of $\Phi_{it}$, we have used a single discount factor $\delta$ so
that $\Delta=\delta I_{dp+1}=\delta I_{2d+1}$. As $\delta$ gets
small (here $\delta=0.8$), the performance of the models
deteriorates. The largest log-likelihood function (for the given
data) is obtained for the model with $d=2$, $\delta=0.98$,
$\beta=0.9$, and it is indicated with boldface in the table. This
model produces also decent values in the $MSSE(1)$, being close to
1. It is worth mentioning that large values of the order of the
TV-VAR $d$ results in models that can approximate time-varying
vector moving average models, e.g. here $d=50$. However, according
to Table \ref{table1} all models with $d=50$ are inferior to that
model with $d=2$, $\delta=0.98$ and $\beta=0.9$. For $\delta=1$ we
obtain a time-invariant VAR model with stochastic volatility, but
this is inferior to the TV-VAR models, even compared with the less
favourable TV-VAR models using $\delta=0.8$ (results not shown
here). It turns out that a slow evolution of the AR matrices
$\Phi_{it}$ (corresponding to $\delta=0.98$) produces the best
results.

\begin{table}
\caption{Log-likelihood function $(\ell(.))$ (evaluated at the
posterior mean of the volatility), mean squared standard one-step
forecast error ($MSSE(1)$) and mean absolute one-step forecast error
($MAE(1)$), for several values of the parameters $d$, $\delta$ and
$\beta$ of the TV-VAR model.}\label{table1}
\begin{center}
\begin{tabular}{|c|rr|rrr|}
\hline & & & $\ell(.)$ & $MSSE(1)$ & $MAE(1)$ \\
\hline $d=1$ & $\delta=0.98$ & $\beta=0.99$ & $-227819.1$ & $[2.408~
2.046]'$
& $[7.375~ 6.365]'$ \\ & & $\beta=0.9$  & $-43000.95$ & $[0.979~0.949]'$ & $[7.375~6.365]'$ \\
& $\delta=0.8$ & $\beta=0.99$  & $-2526833634$ & $[0.498~0.294]'$ &
$[11.798~9.135]'$ \\ & & $\beta=0.9$  &
$-1.352\times 10^{40}$ & $[0.044~0.026]'$ & $[11.798~9.135]'$ \\
\hline $d=2$ & $\delta=0.98$ & $\beta=0.99$ & $-108421.0$ &
$[1.970~1.594]'$ &
$[9.061~7.002]'$ \\ & & $\beta=0.9$ & $\textbf{-12614.11}$ & $[\textbf{0.991}~\textbf{0.975}]'$ &
$[\textbf{9.061}~\textbf{7.002}]'$ \\
 & $\delta=0.8$ & $\beta=0.99$ & $-2212729203$ &
$[0.478~0.131]'$ & $[9.344~7.273]'$ \\ & & $\beta=0.9$ &
$-2.103\times 10^{40}$ & $[0.039~0.013]'$ & $[9.344~7.273]'$ \\
\hline $d=3$ & $\delta=0.98$ & $\beta=0.99$ & $-157643.3$ &
$[1.549~1.388]'$ & $[8.280~6.369]'$ \\ & & $\beta=0.9$  &
$-32741.49$ & $[0.859~0.866]'$ & $[8.280~6.369]'$
\\ & $\delta=0.8$ & $\beta=0.99$ & $-4183496115$ & $[0.094~0.042]'$ & $[8.350~6.426]'$ \\
& & $\beta=0.9$ & $-3.042\times 10^{40}$ & $[0.010~0.006]'$ &
$[8.350~6.426]'$ \\ \hline $d=5$ & $\delta=0.98$ & $\beta=0.99$ &
$-282064.2$ & $[1.445~1.292]'$ & $[8.117~6.236]'$ \\ & & $\beta=0.9$
& $-109868.8$ & $[0.841~0.843]'$ & $[8.117~6.236]'$
\\ & $\delta=0.8$ & $\beta=0.99$ & $-4700987042$ &
$[0.057~0.023]'$ & $[8.146~6.258]'$ \\ & & $\beta=0.9$ &
$-2.032\times 10^{40}$ & $[0.007~0.003]'$ & $[8.146~6.258]'$ \\
\hline $d=10$ & $\delta=0.98$ & $\beta=0.99$ & $-603785.9$ &
$[1.328~1.199]'$ & $[7.704~5.886]'$ \\ & & $\beta=0.9$ & $-306464.6$
& $[0.824~0.826]'$ & $[7.704~5.886]'$
\\ & $\delta=0.8$ & $\beta=0.99$ &
 $-4162987498$ & $[0.049~0.009]'$ & $[7.715~5.893]'$ \\ & &
$\beta=0.9$ & $-1.730\times 10^{40}$ & $[0.005~0.001]'$ &
$[7.715~5.893]'$ \\ \hline $d=50$ & $\delta=0.98$ & $\beta=0.99$ &
$-2786597$ & $[1.131~0.993]'$ & $[7.009~5.733]'$
\\ & & $\beta=0.9$ & $-1796252$ & $[0.814~0.796]'$ &
$[7.009~5.733]'$ \\ & $\delta=0.8$ & $\beta=0.99$ & $-3327435279$ &
$[0.005~0.007]'$ & $[7.010~5.734]'$ \\ & & $\beta=0.9$ &
$-5.694\times 10^{38}$
& $[0.007~0.001]'$ & $[7.010~5.734]'$  \\
\hline
\end{tabular}
\end{center}
\end{table}

\begin{figure}[t]
 \epsfig{file=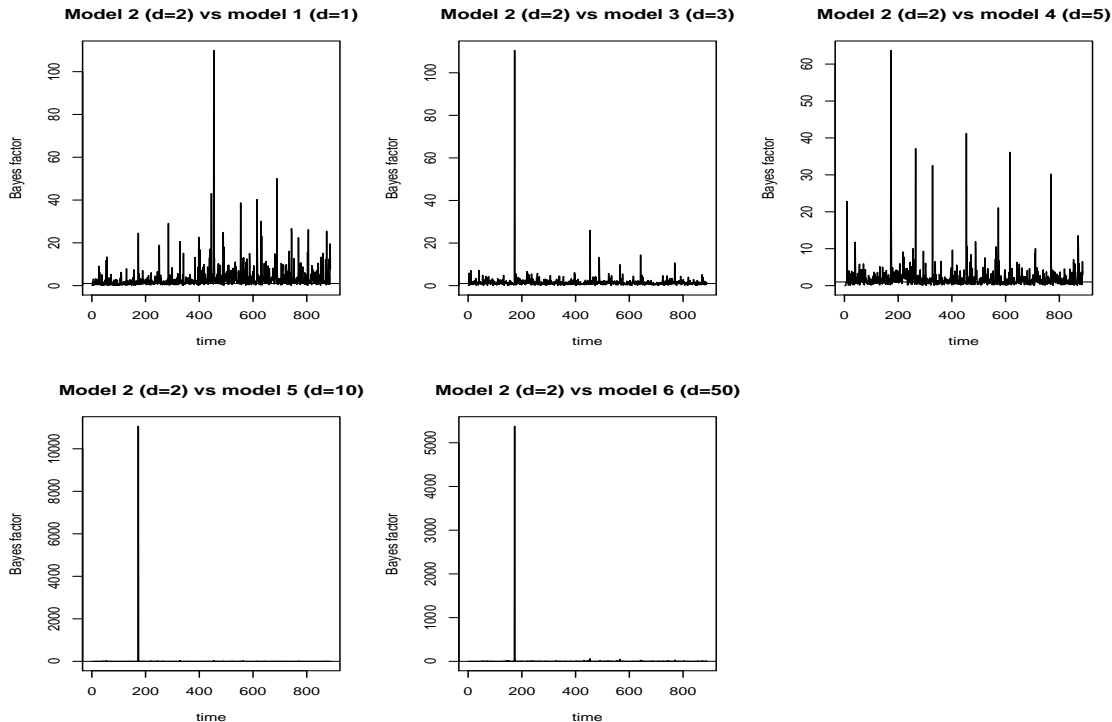, height=10cm, width=15cm}
 \caption{Bayes factor for model 2 ($d=2$) against models 1,3,4,5,6
 ($d=1,3,5,10,50$ respectively), all 6 models having $\delta=0.98,\beta=0.9$.}\label{fig2}
\end{figure}

In order to further compare the chosen model with other models, we
use the Bayes factors as discussed in the previous section. We
compare model 2 ($d=2$) with models 1,3,4,5,6 having respectively
$d=1,3,5,10,50$, where for all 6 models $\delta=0.98$ and
$\beta=0.9$. Figure \ref{fig2} shows 5 plots of the Bayes factors of
model 2 against the other 5 models. From these plots the superiority
of model 2 is clear; for most of the time points model 2 produces
Bayes factors larger than 1. For each of the comparisons the mean of
the Bayeas factor is larger than 1, i.e. 3.487 (model 2 vs model 1),
1.688 (model 2 vs model 3), 2.083 (model 2 vs model 4), 14.965
(model 2 vs model 5), and 9.021 (model 2 vs model 6).

\begin{figure}[t]
 \epsfig{file=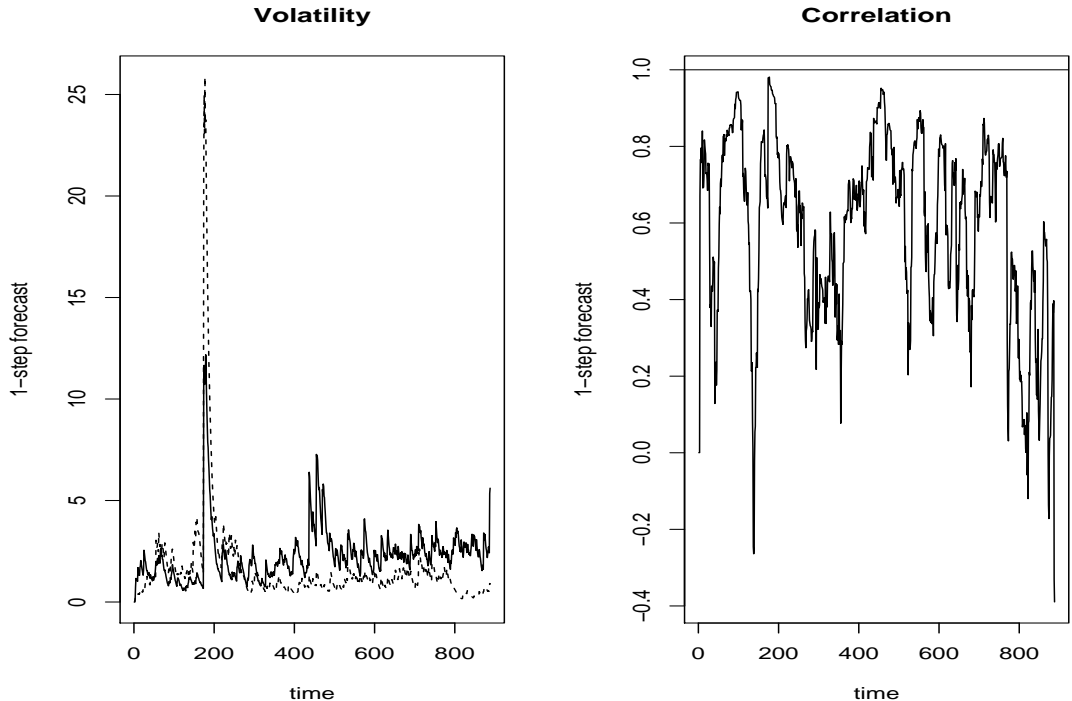, height=10cm, width=15cm}
 \caption{1-step forecasts of the volatility (solid line for the IBM returns and
 dotted line for the S\&P 500 index) and 1-step forecast for the correlation of
 the IBM returns and the S\&P 500 index.}\label{fig3}
\end{figure}

Figure \ref{fig3} shows the one-step forecast of the volatility and
the one-step forecast of the correlation between the IBM shares and
the S\&P 500 index. The forecast of the correlation indicates that
IBM shares are highly correlated with values of the S\&P 500 index
and that this correlation is dynamic.

\begin{figure}
 \epsfig{file=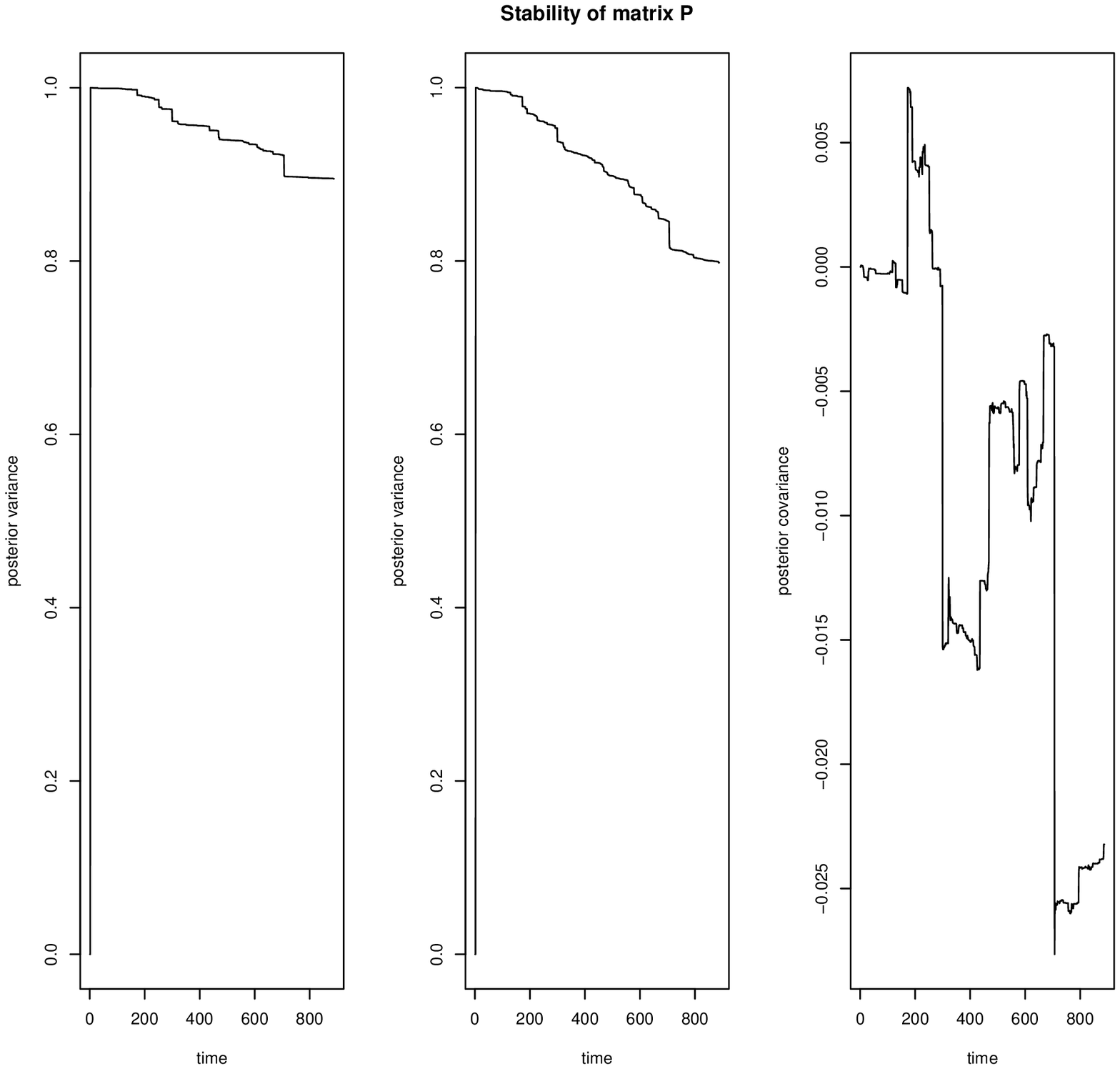, height=8cm, width=15cm}
 \caption{Stability of matrix $\{P_t\}=\{p_{ij,t}\}_{i,j=1,2}$; shown are $\{p_{11,t}\}$ (left panel),
 $\{p_{22,t}\}$ (mid panel) and $\{p_{12,t}\}$ (right panel).}\label{fig4}
\end{figure}

In the previous section it is shown that the sequence of the
matrices $\{P_t\}$ is bounded and this was a key argument on the
stability of the forecast covariance matrix. Figure \ref{fig4} plots
the values of $\{P_t\}$, from which we can see that the diagonal
elements of $\{P_t\}$ are bounded above by 1, while the off-diagonal
elements of $\{P_t\}$ are bounded below by -0.030 and above by
0.007. However, we should note that, although $\{P_t\}$ is bounded,
it does not converge to a stable matrix $P$, something that is usual
in dynamic linear models with time-invariant design components.

One of the advantages of the proposed model is its capability to
forecast future values of the time series vector, but also to
forecast the volatility. Table \ref{table2} shows three performance
measures of multi-step forecasting, ranging from $h=1$ (one day
forecast) to $h=20$ (twenty day forecast). As expected the forecast
performance deteriorates as $h$ increases, but the model is still
usable for a 5-day forecast (corresponding to a weekly forecast,
considering only trading days). The 20-day forecast (corresponding
to a monthly forecast) produces very large $MAE(1)$ and $ME(1)$. It
is interesting to note that throughout the range of $h$, the
$MSSE(1)$ retains reasonable values (being close to 1), where for
$h\geq 3$ the forecast covariance matrix is slightly underestimated.

\begin{table}
\caption{Three performance measures ($MSSE(h)$, $MAE(h)$ and
$ME(h)$) for multi-step forecasting (for several values of the
forecast horizon $h$) using the TV-VAR model
($d=2,\delta=0.98,\beta=0.9$), for the IBM data.}\label{table2}
\begin{center}
\begin{tabular}{|r|rrr|}
\hline & $MSSE(h)$ & $MAE(h)$ & $ME(h)$ \\ \hline $h=1$ &
$[0.991~0.975]'$ & $[9.061~7.002]'$ & $[0.224~0.378]'$ \\ $h=2$ &
$[0.949~0.962]'$ & $[11.500~9.763]'$ & $[0.205~1.458]'$ \\ $h=3$ &
$[1.128~1.084]'$ & $[20.940~20.614]'$ & $[2.681~7.323]'$ \\ $h=4$ &
$[1.468~1.221]'$ & $[59.796~73.719]'$ & $[24.114~43.938]'$ \\ $h=5$
& $[1.360~1.481]'$ & $[259.919~369.380]'$ & $[170.750~294.774]'$ \\
$h=10$ & $[2.034~1.826]'$ & $[1755316~2778512]'$ &
$[1628008~2657889]'$ \\ $h=15$ & $[1.828~2.265]'$ &
$[14641429303~23426293367]'$ & $[14316186757~23117364799]'$ \\
$h=20$ & $[2.785~2.919]'$ & $[1.243997\times 10^{14}~1.996699\times
10^{14}]'$ & $[1.235505\times 10^{14}~1.988631\times 10^{14}]'$ \\
\hline
\end{tabular}
\end{center}
\end{table}

\section{Sequential portfolio allocation: foreign exchange
data}\label{example:fx}

\begin{figure}[t]
 \epsfig{file=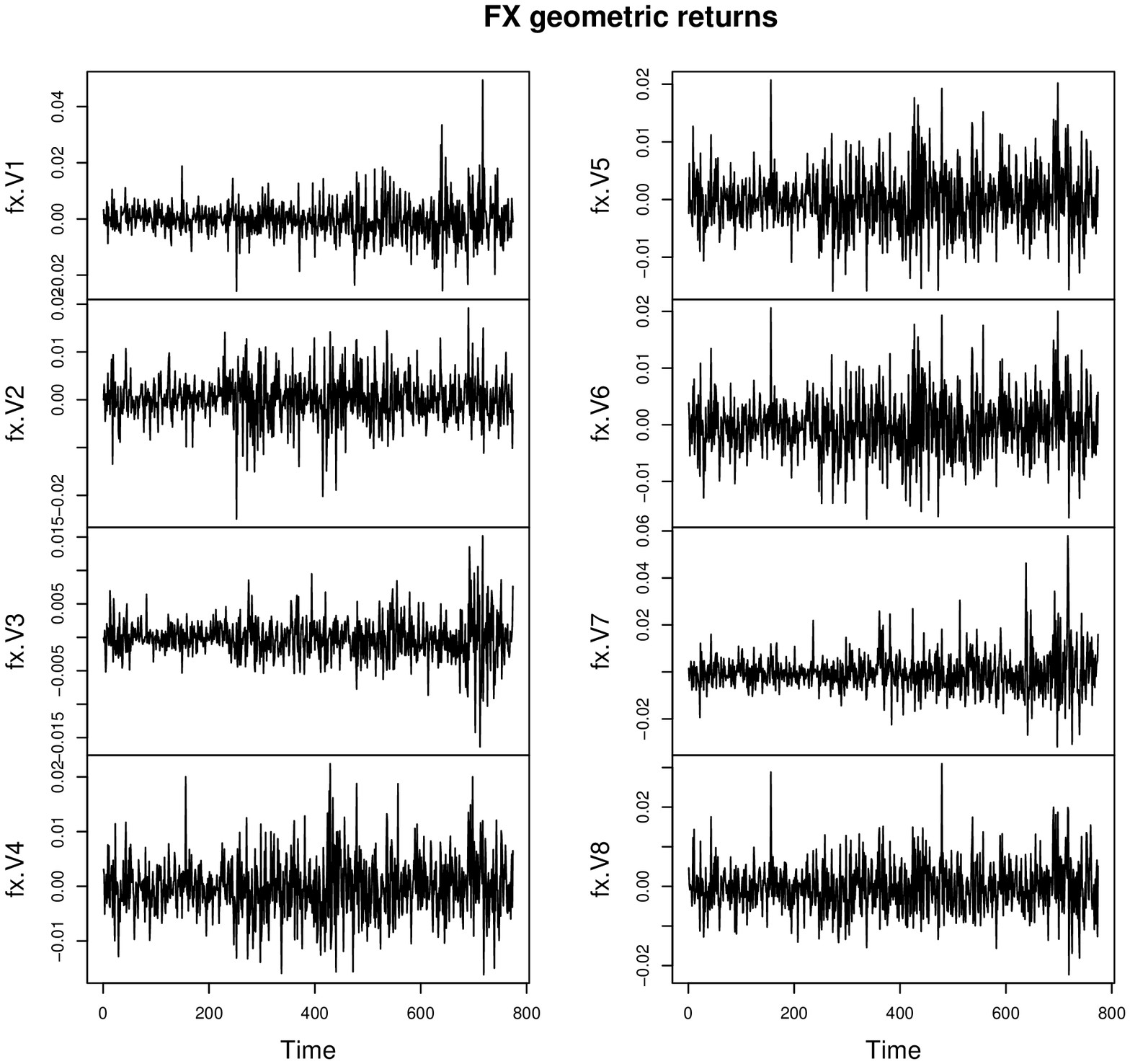, height=10cm, width=15cm}
 \caption{FX data; shown are the geometric returns $y_{it}=x_{it}/x_{i,t-1}-1$
 of each of the currency $i=1,\ldots,8$.}\label{fig5}
\end{figure}

We consider optimal portfolio allocation for FX data. The exchange
rates are the Australian dollar (AUD), British pounds (GBP),
Canadian dollar (CAD), German Deutschmark (GDM), Dutch guilder
(DUG), French frank (FRF), Japanese yen (JPY) and Swiss franc (SWF),
all expressed as number of units of the foreign currency per US
dollar. The data, part of which are discussed in Franses and Dijk
(2000), consist of 774 daily observations from 2 January 1995 to 31
December 1997. Let $x_{it}$ be the exchange rate of currency $i$
$(i=1,\ldots,8)$ and $y_t=[y_{1t}~\cdots~y_{8t}]'$ be the
8-dimensional time series vector ($t=1,\ldots,773$) consisting of
the geometric returns $y_{it}=x_{it}/x_{i,t-1}-1$ (shown in Figure
\ref{fig5}) of the above exchange rates; $y_{1t}$ being the return
of AUD, $\ldots$, $y_{8t}$ being the return of SWF. Here, as in
Aguilar and West (2000), we have chosen to use geometric returns for
the analysis, but log-returns can also be used (Soyer and Tanyeri,
2006).

Sequential portfolio allocation aims to find at each time $t$ an
optimal allocation vector $a_t$, including the proportions of each
currency to be invested, so that the expected return $a_t'f_t=m$
is constant, where $f_t=y_{t-1}(1)$ is the one-step forecast mean
of $y_t$. Assuming no transaction costs and free reallocation of
US dollars over all currencies, the realized return $r_t=a_t'y_t$
can be used to judge the performance of the weights $a_t$. Several
portfolio allocation rules can be considered for the evaluation of
$a_t$, but here we adopt a sequential version of the Markowitz
mean-variance optimization allocation, according to which $a_t$ is
chosen to minimize the variance of $r_t|y^{t-1}$, that is minimize
$\var(r_t|y^{t-1})=a_t'Q_ta_t$, where $Q_t=Q_{t-1}(1)$ is the
one-step forecast covariance matrix of $y_t|y^{t-1}$. It is well
known that the above allocation maximizes the expected return
$a_t'f_t$ (if this is not set to be equal to $m$) given that the
variance of $r_t$ is constant, i.e. $a_t'Q_ta_t=\sigma^2$, for
some $\sigma^2>0$. The unconstrained portfolio (UP) strategy
computes the optimal weights as
$$
a_t=\frac{mQ_t^{-1}f_t}{f_t'Q_t^{-1}f_t}.
$$

The constrained portfolio (CP) allocation rule further constrains
$a_t$ to satisfy $a_t'1_p=1$, where $1_p=[1~\cdots~1]'$, and then
effectively solves a constrained quadratic optimization problem.
The solution is
$$
a_t=Q_t^{-1}\left( \frac{1_p'
Q_t^{-1}(1_pm-f_t)f_t-f_t'Q_t^{-1}(1_pm-f_t)1_p}{
(1_p'Q_t^{-1}1_p)(f_t'Q_t^{-1}f_t)-(1_p'Q_t^{-1}f_t)^2}\right).
$$

Finally the naive equal weight portfolio (EWP) sets
$a_t=[1/p~\cdots~1/p]$ so that each $y_{it}$ is allocated the same
weight $1/p$, where $y_t=[y_{1t}~\cdots~y_{pt}]'$ and $p\geq 2$.
Similar portfolio allocation strategies are discussed in Aguilar
and West (2000), in Soyer and Tanyeri (2006) and in references
therein.

Table \ref{table3} shows the mean of the cumulative realized returns
(\%) for several time series models. We use TV-VAR models with
$m=0.001$, order $d=1,5,10,15$, $\delta=0.96,1$ and $\beta=0.97$
(here as in the IBM example we use a single discount factor
$\delta$, i.e. $\Delta=\delta I_{dp+1}$). $m=0.001$ has been chosen
from historical returns; a similar approach is adopted in Aguilar
and West (2000). First we observe that, as expected, the EWP
realizes loss throughout the models considered. Under the CP, the
TV-VAR outperforms the VAR model, expect for $d=15$. Under the UP,
TV-VAR does again better (except for $d=5$). We also note that under
CP, all VAR models $(\delta=1)$ return loss. It is clear that the
TV-VAR model with $d=10$ and $\delta=0.96$ produces the best results
with both CP and UP being superior compared to the other models.
Models with larger orders tend to be increasingly slow having a
large amount of parameters to be estimated.

\begin{table}
\caption{Mean of cumulative portfolio returns (\%) under three
allocation rules: constraint portfolio (CP), unconstraint
portfolio (UP) and equal weight portfolio (EWP). A TV-VAR model is
used with order $d$, discount factor $\delta$, $\beta=0.97$ and
$m=0.001$.}\label{table3}
\begin{center}
\begin{tabular}{|r|rr|rr|r|}
\hline & CP $(\delta=0.96)$ & UP $(\delta=0.96)$ & CP $(\delta=1)$
& UP $(\delta=1)$ & EWP
\\ \hline $d=1$ & 7.459 & 0.294 & -3.814 & -0.014 & -8.679  \\
$d=5$ & 13.759 & 0.073 & -3.910 & 0.100 & -8.464 \\ $d=10$ & 36.918
& 0.460 & -7.248 & 0.118 & -8.602
\\ $d=15$ & -20.607 & 0.111 & -7.063 & 0.083 & -8.848 \\
\hline
\end{tabular}
\end{center}
\end{table}

\begin{figure}[t]
 \epsfig{file=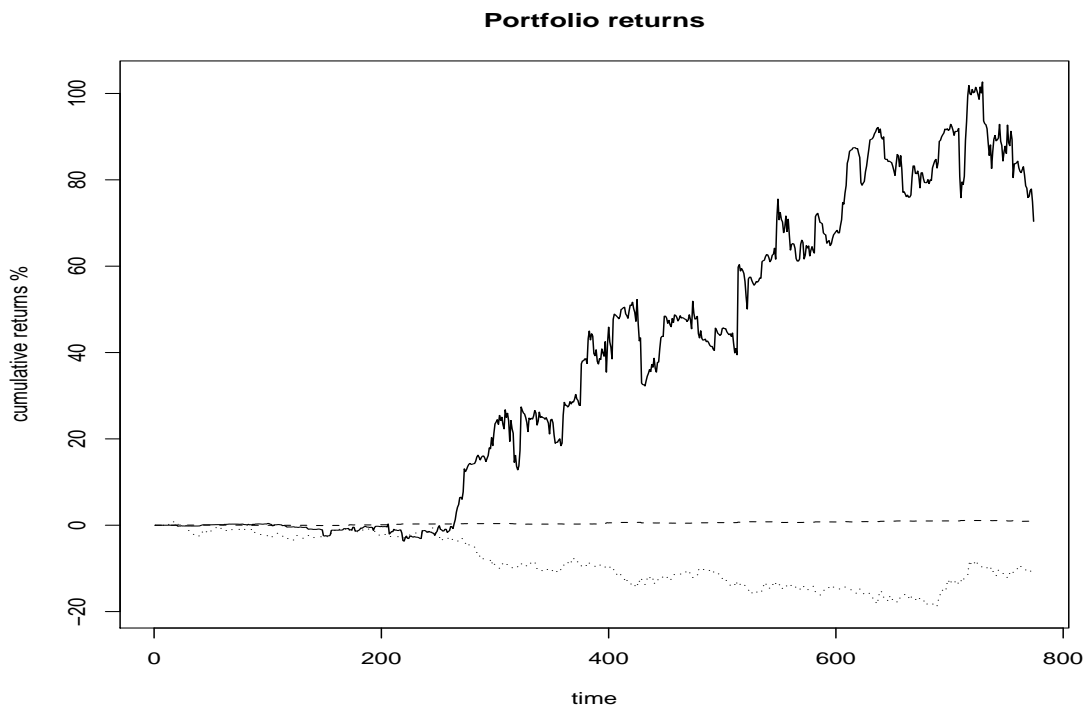, height=10cm, width=15cm}
 \caption{Portfolio allocation for the 8-variate time series of FX
 data; shown are the \% cumulative returns using the constrained portfolio (solid line),
 the unconstrained portfolio (dashed line) and the equal weight portfolio (dotted line).}\label{fig6}
\end{figure}

\begin{figure}[t]
 \epsfig{file=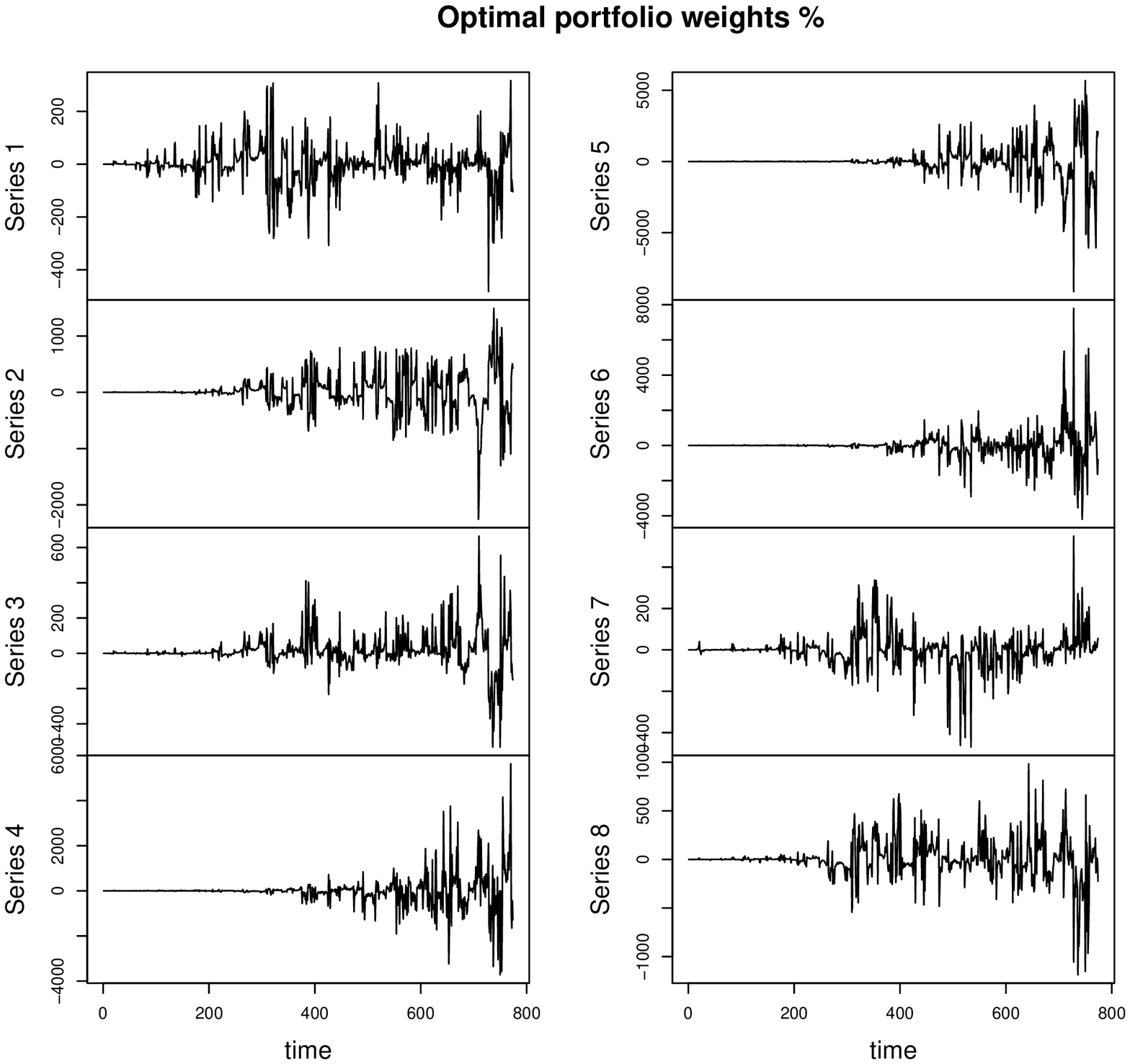, height=10cm, width=15cm}
 \caption{Portfolio allocation for the 8-variate time series of FX data.
 Shown are the \% optimal weights of 8 currencies (AUD, GBP, CAD, DUG,
 FRF, GDM, JPY, and SWF).}\label{fig7}
\end{figure}

For the chosen model with $d=10$, Figure \ref{fig6} shows the
percentage realized cumulative returns. From this graph we observe a
high realized portfolio profit using the CP; the UP achieves
marginal profit, and the EWP realizes significant loss. The
corresponding optimal portfolio weights $a_t$ (for the CP) are shown
in Figure \ref{fig7}.

\section{Concluding comments}\label{conclusions}

In this paper we have proposed the time-varying vector
autoregressive model (TV-VAR) for modelling multivariate time
series. Adopting Bayesian inference, we propose the use of conjugate
priors, resulting in fast and accurate computations. Doubt in the
choice of the hyperparameters of the priors, can be eliminated by
using relatively long time series data, which are usually available
in financial time series. The model includes an important element of
stochastic volatility, which makes the model realistic and very
useful, as most of financial time series exhibit heteroscedasticity.
The volatility is estimated using inverted Wishart and singular
multivariate beta distributions, which form a fast and flexible
multiplicative evolutionary law for the volatility. The proposed
approach offers the advantage of combining in one model forecasting
of the time series vector as well as forecasting of the volatility
covariance matrix. Using bivariate data of IBM shares and FX data of
8 currencies we provide empirical evidence in favour of the TV-VAR.
Our empirical findings suggest that TV-VAR models work better than
the usual VAR models. We believe that time-varying models offer
important advantages for financial time series forecasting and
future research is likely to be devoted in this direction.

\section*{Acknowledgements}

I should like to thank Nick Firoozye for helpful comments on an
earlier version of the paper.


\begin{thebibliography}{}

\bibitem{Abramovich07}
Abramovich, Y.I., Spencer, N.K. and Turley, M.D.E. (2007) Order
estimation and discrimination between stationary and time-varying
(TVAR) autoregressive models. {\it IEEE Trnasactions on Signal
Processing}, {\bf 55}, 2861-2876.

\bibitem{Aguilar00}
Aguilar, O. and West, M. (2000) Bayesian dynamic factor models and
portfolio allocation. {\it Journal of Business and Economic
Statistics}, {\bf 18}, 338-357.

\bibitem{Andrieu03}
Andrieu, C., Davy, M. and Doucet, A. (2003) Efficient particle
filtering for jump Markov systems. Application to time-varying
autoregressions. {\it IEEE Transactions on Signal Processing}, {\bf
51}, 1762-1770.

\bibitem{Dahlhaus}
Dahlhaus, R. (1997) Fitting time series models to nonstationary
processes. {\it Annals of Statistics}, {\bf 25}, 1-37.

\bibitem{Waele03}
de Waele, S. and Broersen, M.T. (2003) Order selection for vector
autoregressive models. {\it IEEE Transactions on Signal Processing},
{\bf 51}, 427-433.

\bibitem{Franses00}
Franses, P.H. and Dijk, D. (2000) {\it Nonlinear Time Series Models
in Empirical Finance.} Cambridge University Press, Cambridge.

\bibitem{Francq}
Francq, C. and Gautier, A. (2004) Large sample properties of
parameter least squares estimates for time-varying ARMA models. {\it
Journal of Time Series Analysis}, {\bf 25}, 765-783.

\bibitem{Hamilton}
Hamilton, J.D. (1995) {\it Time Series Analysis.} Princeton
University Press, Princeton.

\bibitem{Huerta}
Huerta, G. and Prado, R. (2006) Structured priors for multivariate
time series. {\it Journal of Statistical Planning and Inference},
{\bf 136}, 3802-3821.

\bibitem{Jiang}
Jiang, X.Q. and Kitagawa, G. (1993) A time-varying coefficient
vector AR modelling of nonstationary covariance time-series. {\it
Signal Processing}, {\bf 33}, 315-331.

\bibitem{Kitagawa85}
Kitagawa, G. and Gersch, W. (1985) A smoothness priors time varying
AR coefficient modeling of nonstationary time series. {\it IEEE
Transanctions in Automatic Control}, {\bf AC-30}, 48-56.

\bibitem{Kitagawa96}
Kitagawa, G. and Gersch, W. (1996) {\it Smoothness Priors Analysis
of Time Series,} Lecture Notes in Statistics, vol. 116,
Springer-Verlag, New York.

\bibitem{Lundbergh03}
Lundbergh, S., Terasvirta, T. and van Dijk, D. (2003) Time-varying
smooth transition autoregressive models. {\it Journal of Business
and Economic Statistics}, {\bf 21}, 104-121.

\bibitem{Lutkepohl05}
L\"utkepohl, H. (2005) {\it New Introduction to Multiple Time Series
Analysis}. Springer-Verlag, Berlin.

\bibitem{Moulines}
Moulines, E., Priouret, P. and Roueff, F. (2005) On recursive
estimation of time varying autoregressive processes. {\it Annals of
Statistics}, {\bf 33}, 2610-2654.

\bibitem{Ni03}
Ni, S. and Sun, D. (2003) Noninformative priors and frequentist
risks of Bayesian estimators of vector-autoregressive models. {\it
Journal of Econometrics}, {\bf 115}, 159-197.

\bibitem{Prado}
Prado, R. and Huerta, G. (2002) Time-varying autoregressions with
model order uncertainty. {\it Journal of Time Series Analysis}, {\bf
23}, 599-618.

\bibitem{Sarantis06}
Sarantis, N. (2006) On the short-term predictability of exchange
rates: A BVAR time-varying parameters approach. {\it Journal of
Banking and Finance}, {\bf 30}, 2257-2279.

\bibitem{Sato07}
Sato, J.R., Morettin, P.A., Arantes, P.R. and Amaro, E. (2007)
Wavelet based time-varying vector autoregressive modelling. {\it
Computational Statistics and Data Analysis}, {\bf 51}, 5847-5866.

\bibitem{Sims80}
Sims, C.A. (1980) Macroeconomics and reality. {\it Econometrica},
{\bf 48}, 1-48.

\bibitem{Soyer06}
Soyer, R. and Tanyeri, K. (2006) Bayesian portfolio selection with
multi-variate random variance models. {\it European Journal of
Operational Research}, {\bf 171}, 977–990.

\bibitem{Tiao}
Tiao, G.C. and Tsay, R.S. (1989) Model specification in multivariate
time series. {\it Journal of the Royal Statistical Society Series
B}, {\bf 51}, 157-213.

\bibitem{trianta07}
Triantafyllopoulos, K. (2007) Covariance estimation for multivariate
conditionally Gaussian dynamic linear models. {\it Journal of
Forecasting}, {\bf 26}, 551-569.

\bibitem{triantanason}
Triantafyllopoulos, K. and Nason, G.P. (2007) A Bayesian analysis of
moving average processes with time-varying parameters. {\it
Computational Statistics and Data Analysis}, {\bf 52}, 1025-1046.

\bibitem{trianta08}
Triantafyllopoulos, K. (2008) Multivariate stochastic volatility
with Bayesian dynamic linear models. {\it Journal of Statistical
Planning and Inference}, {\bf 138}, 1021-1037.

\bibitem{Tsay02}
Tsay, R.S. (2002). {\it Analysis of Financial Time Series}. Wiley,
New York.

\bibitem{Uhlig94}
Uhlig, H. (1994) On singular Wishart and singular multivariate beta
distributions. {\it Annals of Statistics}, {\bf{22}}, 395-405.

\bibitem{Uhlig97}
Uhlig, H. (1997) Bayesian vector autoregressions with stochastic
volatility. {\it Econometrica}, {\bf{65}}, 59-73.

\bibitem{West97}
West, M. and Harrison, P.J. (1997). {\it Bayesian Forecasting and
Dynamic Models}. Springer-Verlag, 2nd edn., New York.

\bibitem{West99}
West, M., Prado, R. and Krystal, A.D. (1999) Evaluation and
comparison of EEG traces: latent structure in nonstationary time
series. {\it Journal of the American Statistical Association}, {\bf
94}, 375-387.

\bibitem{Zhu07}
Zhu, H. and Wu, H. (2007) Estimation of smooth time-varying
parameters in state space models. {\it Journal of Computational and
Graphical Statistics}, {\bf 16}, 813–-832.

\end{thebibliography}
\end{document}